\documentclass[12pt]{iopart}

\usepackage[dvips]{graphicx}
\usepackage{iopams}

\usepackage{bbold}
\usepackage{dsfont}
\usepackage{textcomp}
\usepackage{upgreek}
\usepackage{color}

\DeclareMathAlphabet{\mathpzc}{OT1}{pzc}{m}{it}

\def\Rb87{^{87}\text{Rb}} 
\def\Na23{^{23}\text{Na}} 
\def\Li6{^{6}\text{Li}} 

{\catcode`\|=\active
 \gdef\Braket#1{\left<\mathcode`\|"8000\let|\BraVert {#1}\right>}}
\def\BraVert{\egroup\,\mid@vertical\,\bgroup}

\begin{document}

\title{Direct observation of zitterbewegung in a Bose-Einstein condensate}
\author{L J LeBlanc$^1$
, 
M C Beeler$^1$\footnote{Present address: The Johns Hopkins Applied Physics Laboratory, Laurel, MD 20723, USA}, 
K Jim{\'e}nez-Garc{\'i}a$^{1,2}$\footnote{Present address: The James Franck Institute and Department of Physics, The University of Chicago, Chicago, IL 60637, USA},
 A R Perry$^1$, 
 S Sugawa$^1$,
  R A Williams$^1$\footnote{National Physical Laboratory, Teddington TW11 0LW, UK} and
  I B Spielman$^1$}

\address{$^1$Joint Quantum Institute, National Institute of Standards and Technology, and University of Maryland, Gaithersburg, Maryland, 20899, USA}
\address{$^2$Departamento de F\'{\i}sica, Centro de Investigaci\'{o}n y Estudios Avanzados del Instituto Polit\'{e}cnico Nacional, M\'{e}xico D.F., 07360, M\'{e}xico}

\ead{\mailto{ian.spielman@nist.gov}}

\begin{abstract}
Zitterbewegung, a force-free trembling motion first predicted for relativistic fermions like electrons, was an unexpected consequence of the Dirac equation's unification of quantum mechanics and special relativity.  Though the oscillatory motion's large frequency and small amplitude have precluded its measurement with electrons, zitterbewegung is observable via quantum simulation.  We engineered an environment for $^{\bf 87}$Rb Bose-Einstein condensates where the constituent atoms behaved like relativistic particles subject to the one-dimensional Dirac equation.  With direct imaging, we observed the sub-micrometer trembling motion of these clouds,  demonstrating the utility of neutral ultracold quantum gases for simulating Dirac particles. 
\end{abstract}

\pacs{67.85.Hj, 03.67.Lx, 03.65.Pm}

\maketitle
\section{Introduction}
Among the great discoveries of the Enlightenment was the realization that physical laws are equivalent in all places, at all times, and for all scales; this remains a central tenet in contemporary science.  Quantum simulation exploits this universality to study the behaviour of systems that are difficult to access or impossible to manipulate, by performing direct measurements on analogue systems composed of well-characterized and highly manipulable quantum building blocks.  In this work, we used neutral rubidium atoms to simulate zitterbewegung, a trembling motion usually associated with relativistic electrons~\cite{Schroedinger1930}, and we illuminate its microscopic origins by drawing an analogy to the well-understood atomic physics of Rabi oscillations.  The Dirac equation -- describing the motion of free fermions -- is an essential part of our current description of nature; by engineering new Dirac particles in novel settings, we expose the equation's properties by direct measurement. Simulations of the Dirac equation have been proposed for superconductors~\cite{Lurie1970}, semiconductors~\cite{Schliemann2005,Zawadzki2011}, graphene~\cite{Katsnelson2006}, cold atoms~\cite{Zawadzki2011,RuostekoskiPRL2002,Vaishnav2008,Juzeliunas2008,Merkl2008,Larson2010,Zhang2010,Salger2011,Zhang2012a,Zhang2012}, and photonic systems~\cite{Zhang2008b}; and have been realized with cold atoms~\cite{KlingPRL2010}, trapped ions~\cite{Gerritsma2010}, and photons~\cite{DreisowPRL2010}.   
The ion and photon experiments demonstrated zitterbewegung for quantities analogous to position or time in the Dirac equation.  
Here, we directly observed a neutral-atom BEC undergoing zitterbewegung in space and time.  

Zitterbewegung, as observed here, is an example of a broader class of phenomena where a group of states with differing velocities are quantum mechanically coupled together and undergo Rabi-like oscillations~\cite{MellishPRA2003,Katz2004,David2010,Qu2013}.  As with the present case, eigenstates of the full Hamiltonian are static, but superpositions can tremble.  Neutrino oscillations~\cite{Gonzalez-Garcia2003} are an example of this generalization: neutrinos are produced by the weak nuclear force in superpositions of the propagating (i.e., mass) eigenstates, each with a different mass and, therefore, velocity.

\begin{figure}[h!]
\begin{center}
\includegraphics{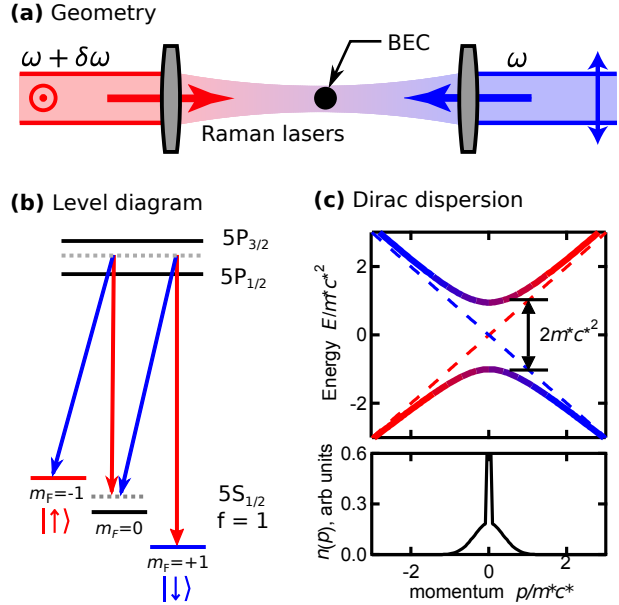}
\end{center}
\caption{(a) Schematic laser geometry.  Two counter propagating laser beams (red and blue) coupled the Zeeman levels of the $^{87}{\rm Rb}$ BEC's $f=1$ ground state. (b) Coupled states of $^{87}$Rb. The $\left|\uparrow\downarrow\right> = \left|m_F = \mp1\right>$ states were laser coupled with a four-photon Raman transition, connecting states differing in velocity by $4 v_{\rm R}$. (c)  Top panel: Dirac dispersion relationship for massless particles (dashed lines) and for massive particles such as electrons and positrons (solid curves).   With suitable values of $m^*$ and $c^*$, this same dispersion relationship and its underlying Dirac Hamiltonian equally describe relativistic electrons and our atomic system. In the vicinity of the depicted avoided crossing, atoms in $\left|\uparrow\right>$ move with velocities near $2 v_{\rm R}$, and those in $\left|\downarrow\right>$ have velocities near $-2 v_{\rm R}$.  Bottom panel: Typical momentum distribution of the BEC (narrow peak) and thermal cloud (broad) in our system.  The vertical axis is truncated to show detail -- the central peak reaches a value of 18 on this scale.}
\label{fig:zitter1}
\end{figure}

The precise control and direct measurement techniques available in systems of ultracold atoms, coupled with their accessible length and energy scales, make these systems ideal for quantum simulation.  In this experiment, our quantum building blocks were Bose-condensed $^{87}$Rb atoms.  Using two counter-propagating Raman lasers [figure~1(a)] with wavelength $\lambda = 790.1$~nm, we coupled the atoms' $\left | f = 1, m_F = \mp 1\right\rangle = \left|\uparrow\downarrow\right>$ atomic hyperfine states (comprising our effective two level system) to their external motion~\cite{Lin2011} with a four-photon Raman transition [figure~1(b)].  In this environment, each atom's behaviour was governed by the one-dimensional Dirac Hamiltonian, making its motion analogous to that of a relativistic electron. The system's characteristic momentum $\hbar k_{\rm R} = 2\pi\hbar/\lambda$  -- that of a single photon --  specifies the recoil energy $E_{\rm R} = \hbar^2 k_{\rm R}^2/2m = h\times3.68\ {\rm kHz}$, where $m$ is the atomic mass.  These recoil units set the scale for all physical quantities in our analogue system, such as the recoil velocity $v_{\rm R}  = \hbar k_{\rm R}/m$.  The Raman lasers drove the four-photon $\left|\uparrow, \hbar k_x = p_x + 2\hbar k_{\rm R} \right>\leftrightarrow\left|\downarrow, \hbar k_x = p_x - 2\hbar k_{\rm R}\right>$ transition (resonant when $p_x = 0$), where $\hbar k_x$ is the atomic momentum along ${\bf e}_x$ and $p_x$ will play the role of momentum in the Dirac equation.  The simulated speed of light $c^* = 2v_{\rm R} = 11.6$~mm/s was twice the atoms' recoil velocity, a factor of  $\approx10^{10}$ less than the true speed of light. The artificial rest energy $m^{*}{c^*}^2 = \hbar\Omega / 2\lesssim 1~E_{\rm R}$ was a factor of  $\approx10^{17}$ less than the electron's rest energy ($\hbar\Omega$ is the four-photon laser coupling strength).  The effective Compton wavelength $\lambda^*_{\rm C} = h/m^* c^* \approx 1\ \upmu{\rm m}$, the approximate amplitude of zitterbewegung, exceeded that of an electron by a factor of $\approx10^6$.  These new scales enabled our direct measurement of zitterbewegung.

The dynamics of our ultracold $^{87}$Rb atoms were described by the  one-dimensional Dirac equation
\begin{equation}
	\hat H_{\rm D}\left|\psi\right> =  \left(c^* \hat{p}_x \check\sigma_z + m^* {c^*}^2 \check\sigma_x\right)\left|\psi\right>,
	\label{eq:Dirac}
\end{equation}
where $\hat p_x$ is the momentum operator;  $\check\sigma_{x,y,z}$ are the Pauli spin operators; and $\left|\psi\right>$ is represented as a two-component spinor, whose components are defined by $\left|\uparrow\downarrow,p_x\right>$, the $m^*=0$ eigenstates of $\hat H_{\rm D}$.  For the massless, $m^* = 0$ case, this equation simply describes  particles (positive energy) or anti-particles (negative energy) travelling with velocity $\pm c^*$, as depicted by the dashed lines in figure~1(c). The mass term couples together these $m^* = 0$ states, producing an avoided crossing [solid curves in figure~1(c)] with energy given by the familiar relativistic dispersion $E(p_x) = \pm(p_x^2 {c^*}^2 + {m^*}^2{c^*}^4)^{1/2}$, gapped at $p_x = 0$ by twice the rest energy.  In our atomic analogue, the two massless  states coupled by the effective rest energy physically corresponded to the atomic states $\left|\uparrow\downarrow\right>$ moving with  velocity  $\pm c^*$.

\subsection{Zitterbewegung equations of motion}
Zitterbewegung arises because the Pauli matrices associated with the two terms in the Dirac equation do not commute.  In the Heisenberg representation of quantum mechanics the operators, not the wavefunctions, depend on time: for example $\hat v_x = d\hat x/dt = [\hat x, \hat H_{\rm D}]/i\hbar$.  In this formalism, the velocity operator obeys the differential equation
\begin{equation}
\frac{d^2\hat v_x}{dt^2} + \Omega^2 \hat v_x = \frac{2\Omega {c^*}^2}{\hbar} \hat p_x \check\sigma_x.
\end{equation}
For an initial state $\left|\uparrow,p_x = 0\right>$, which gives initial conditions $\langle  \hat v_x \rangle = c^*$ and $\langle  d \hat v_x/dt \rangle = 0$, the expectation values of the position and velocity observables oscillate with the zitterbewegung frequency $\Omega$ according to
\begin{equation}
\langle \hat x (t)\rangle = x(0) + \frac{\lambda^*_{\rm C}}{4\pi} \sin(\Omega t);\quad
\langle  \hat v_x (t)\rangle =  c^* \cos(\Omega t).
\label{eq:speed}
\end{equation}
Initial states with $\langle \hat p_x\rangle\neq0$, or localized wave packets, follow more complex trajectories~\cite{Park2012}.  Zitterbewegung, as usually understood, refers to trembling in position;  an oscillatory velocity is the obvious dual.  In these experiments, we observed the out-of-phase oscillation of these conjugate quantities.

\subsection{The atomic Dirac Hamiltonian}

The one-dimensional Dirac Hamiltonian for a system of $^{87}$Rb atoms can be realized by coupling different spin-momentum states.  The three $m_F$ states comprising the $5S_{1/2}$, $f=1$ electronic ground state manifold are subject to a two-photon Raman process [figure~\ref{fig:zitter1}(a)], and the atomic dynamics along ${\bf e}_x$ are described by the three-level Hamiltonian
\begin{eqnarray}
\hat H^{3\times3} = & \left(\frac{\hbar^2{\hat k_x}^2}{2m}+4E_{\rm R}\right)\check 1_3 +  
\frac{2\hbar^2 k_{\rm R} \hat k_x}{m} \check\sigma_{3,z} +\frac{\hbar\Omega_2}{2}\check\sigma_{3,y}  +(\check 1_3- \check\sigma_{3,z} )\hbar\epsilon,
\end{eqnarray} 
where $\Omega_2$ is the 2-photon Raman coupling strength; $\epsilon$ is the quadratic Zeeman shift that energetically displaces the $m_F = 0$ state; $\check\sigma_{3,z}$ are the generalized Pauli operators for a spin-1 system; and $\check 1_3$ is the $3\times3$ identity.   We concentrate on the avoided crossing at $k_x=0$ between the states that adiabatically connect to $\left|m_F = -1\right>$ and $\left|m_F = +1\right>$. By adiabatically eliminating the lowest-energy eigenstate, we obtain the effective two-level Hamiltonian
\begin{equation}
		 \hat H^{2\times2} = \left[ \frac{\hbar^2 (\hat k_x^2+4k_{\rm R}^2)}{2m} +E_4 \right] \check 1 + 
		 \frac{2 \hbar^2 k_{\rm R} \hat k_x}{m} \check\sigma_{x} +\frac{\hbar\Omega }{2}\check\sigma_{z},
		 \label{eq:Hat}
\end{equation} 
which includes a  global rotation of the system $\check\sigma_x \rightarrow \check\sigma_y$, $\check\sigma_y \rightarrow \check\sigma_z$, $\check\sigma_z \rightarrow \check\sigma_x$.
 For $ k_x/k_{\rm R} \ll 1$, the effective coupling is $\Omega = {\hbar\Omega_2^2}/{2(4E_L-\hbar\epsilon})$ and $E_4 = E_L \Omega/\Omega_2 $.  Ignoring the uniform energy offset, we identify the parameters from the Dirac Hamiltonian (1): the effective  $c^* = 2\hbar k_{\rm R}/m$ is twice the atomic recoil velocity, the rest energy $m^*{c^*}^2 = \hbar \Omega/2$ is the coupling strength, and the Compton wavelength $\lambda^*_{\rm C} = h/m^*c^* = 8\pi \hbar k_{\rm R} / m \Omega$ sets the scale for the zitterbewegung's amplitude.  The equivalence of this Hamiltonian \eref{eq:Hat} and the Dirac Hamiltonian \eref{eq:Dirac} provides the opportunity for our quantum simulation of relativistic electron dynamics.

\section{Experimental techniques}

To study zitterbewegung with an ultracold atomic gas, we measured the positions and velocities of atomic systems subject to the Dirac Hamiltonian for varying times after starting in an initial state with speed $c^*$.   These experiments began with $N\approx5\times10^4$ atom optically-trapped $^{87}$Rb BECs [$f_c=0.75(10)$ condensate fraction] in the $\left|f = 1, m_{F} = -1\right\rangle$ ground state, subject to a uniform $B_0 = 2.1$~mT bias magnetic field.  The atoms were confined in a harmonic trap [$(\omega_x,\omega_y,\omega_z)/2\pi$ = (38,38,130) Hz] with characteristic timescales greatly exceeding those of the zitterbewegung. We transferred these atoms (at rest) to $\left|f = 1, m_{F} = 0\right\rangle$ using an adiabatic rapid passage technique; a fixed frequency 15.0~MHz radiofrequency magnetic field coupled the different $m_F$ states together as the bias magnetic field was swept through resonance.  Using a pair of  Raman beams counterpropagating  along ${\bf e}_x$ with wavelength $\lambda = 790.1$~nm and frequency difference $\delta \omega = g_F \upmu_{\rm B} B_0 + 4 E_{\rm R} + \epsilon$ (where $\epsilon = h\times 32~{\rm kHz}$ is the quadratic Zeeman shift), a 30~$\upmu$s $\pi$-pulse transferred approximately $85 \%$ of  the atoms from $\left|m_{F} = 0, k_x = 0\right\rangle$ to $\left|m_{F} = -1, k_x = 2 k_{\rm R}\right\rangle$ (moving with velocity $v= 2\hbar k_{\rm R}/m = c^*$).  Before the trap appreciably altered their velocity (200~$\upmu$s), we changed the Raman laser's frequency difference to $\delta \omega = g_F \upmu_{\rm B} B_0$, bringing $|m_{F} = -1, k_x = 2 k_{\rm R}\rangle$ and $|m_{F} = +1, k_x = -2 k_{\rm R}\rangle$ into four-photon resonance.  We then suddenly introduced a four-photon Raman coupling between these states [figure~\ref{fig:zitter1}(b)], and allowed the system to evolve under this new Hamiltonian for an evolution time $t$.

Just before transferring the BEC into $\left|m_F=-1,k_x=2k_{\rm R}\right>$, two 6.8 GHz microwave pulses spaced in time by $50\ {\rm ms}$ each out-coupled $\approx 10\%$ of the atoms to the $f=2$ hyperfine manifold.  These atoms  were separately imaged (without repumping on the $f=1$ to $f=2$ transition) leaving the atoms in $f=1$ undisturbed.  These $f=2$ atoms served two purposes: 1) by setting the microwave frequency 2 kHz above (first pulse) and 2 kHz below (second pulse) resonance, we tracked shifts in the bias field that would change our four-photon Raman resonance condition.  Upon analyzing the data, we rejected points where the atom number difference between these two images was greater than two standard deviations from being equal; 2) we determined the BEC's position immediately before each zitterbewegung experiment began, allowing us to cancel shot-to-shot variations in the trap position.  The beginning of the three transfer pulses -- two microwave outcoupling pulses, and the final four-photon Raman pulse -- were each separated in time by 50~ms. As three periods of a 60~Hz cycle, this separation was chosen to reduce magnetic field background fluctuations at the power line frequency, and to facilitate rethermalization between pulses.

\section{Measurement and analysis}
We measured the system either by imaging the atoms immediately following this evolution (to determine the atoms' position) or by releasing the atoms from their trap and simultaneously turning off the Raman lasers, allowing for a short time-of-flight (TOF, with duration $t_{\rm TOF}$) before imaging (to determine the atoms' velocity).   Figure 2 shows the evolution of the signal for several times of flight, and figure~3  shows in situ and after TOF ($t_{\rm TOF} = 550~\upmu$s) measurements at several coupling strengths; the velocity-dominated TOF images clearly show the expected cosinusoidal behaviour.  For in situ measurements, the Raman and trapping beams remained on during the $40~\upmu$s absorption imaging pulses.  For time-of-flight, these were removed during TOF during which time the atoms flew ballistically for $t_{\rm TOF}$ and were subsequently absorption-imaged.  We used high intensity imaging, with intensity $I \approx 3 I_{\rm sat}$ (where $I_{\rm sat}$ is the saturation intensity), that reduced the effective optical depth~\cite{Reinaudi2007} and gave better signal-to-noise in the determination of the clouds' positions.

\begin{figure}[t!]
\begin{center}
\includegraphics{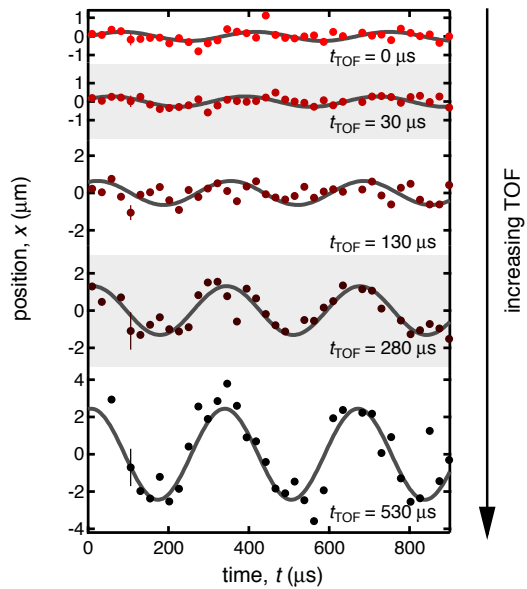}
\end{center}
\caption{Direct detection of zitterbewegung in position and velocity.  We measured the time-evolving position  of the cloud subject to the Dirac Hamiltonian for various times of flight. We plot in situ ($t_{\rm TOF} = 0$)  and TOF data, showing how the data transform from reflecting the atoms' position to indicating their velocity, with  $\Omega = 2\pi\times 3.0$~kHz. 
Statistical uncertainties are shown with typical error bars in each set of data. }
\label{fig:zitter2}
\end{figure}

This simple description of zitterbewegung assumes that the range $\delta p_x$ of occupied momentum states is small compared to $m^* c^*$, and only those states near the avoided-crossing structure are populated.  To maintain a sufficiently narrow $\delta p_x$, the spatial size of the system must be at least $\delta x\gg \lambda^*_{\rm C}$, which, as observed in \cite{OConnell2011}, is larger than the $\lambda^*_C/4\pi$ amplitude of the zitterbewegung itself.  We satisfied this requirement in our experiment by using clouds whose Thomas-Fermi radii $R_x = 12(2)~\upmu{\rm m}$ greatly exceeded the measured sub-micron zitterbewegung oscillations, and overcame the fundamental measurement challenge with good statistics.  Just before initializing zitterbewegung, we measured the initial position of the BEC by out-coupling and imaging $\approx5\times10^3$ atoms.  In principle, this allowed us to measure the centre of the distribution with an uncertainty estimated by $R_x / \sqrt{5\times10^3} \approx 0.17~\upmu{\rm m}$. Our actual measurements, which include technical noise and are averages of four independent images, have a typical $0.3~\upmu{\rm m}$ rms uncertainty [much less than both the distribution's $12(2)~\upmu{\rm m}$ width and our $\approx1.75~\upmu{\rm m}$ imaging resolution].

\begin{figure*}[t!]
\begin{center}
\includegraphics{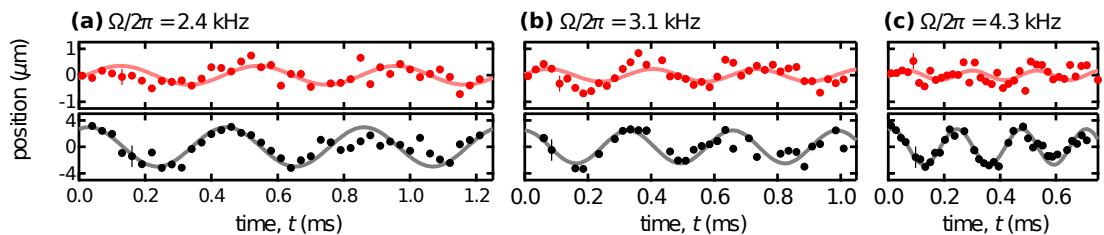}
\end{center}
\caption{Frequency dependence of zitterbewegung in position and velocity.{(a)-(c)} We repeated the in situ (red) and $t_{\rm TOF} = 550~\upmu{\rm s}$ (black) measurements for various coupling strengths, showing the zitterbewegung in position $\left\langle \hat x\right\rangle$ and velocity (where TOF position is proportional to velocity $\left\langle \hat v_x \right\rangle$).  Shown are (a) $\Omega = 2\pi\times 2.4$~kHz, (b) $\Omega = 2\pi\times 3.1$~kHz, and (c) $\Omega = 2\pi\times 4.3$~kHz. 
Statistical uncertainties are shown with typical error bars in each set of data, for five repeated measurements. }
\label{fig:zitter3}
\end{figure*}

\begin{figure}[t!]
\begin{center}
\includegraphics{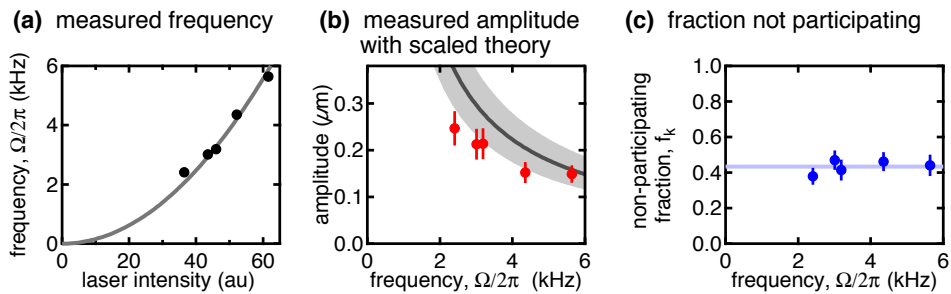}
\end{center}
\caption{ Observed zitterbewegung parameters. (a) Coupling frequency $\Omega/2\pi$ measured for different values of the Raman laser intensity.  As a four-photon transition, this scales quadratically with the laser intensity (grey curve).  Statistical uncertainties are smaller than the symbol size, and the observed scatter likely results from fluctuations in the bias magnetic field.  
 (b) Observed in situ amplitude of zitterbewegung  (symbols), plotted along with our model, including the effects of non-participating atoms with fractions $f_k$ and $f_0 = 15(10)\%$.  Uncertainties in  values are statistical.
 (c) Fraction of atoms $f_k$ not participating in zitterbewegung, and their average (line). Uncertainties are statistical, and derived from the fit that determines the fraction.

}
\label{fig:zitter4}
\end{figure}

From fits to data as in figure~3 -- with parameters joint between each in situ and TOF pair -- we extracted the frequency $\Omega$, amplitude $\lambda^*_C/4\pi$, and velocity $c^*$ of the observed zitterbewegung (shown in figure~4).  The observed values are attenuated by approximately 2.5 from those predicted by (\ref{eq:speed}), as explained below.

\subsection{Amplitude attenuation}
For our finite-temperature system, higher momentum states are thermally occupied in the initial equilibrium system [schematically illustrated in figure~1(c) and observed in figure~5].  The zitterbewegung frequency for these states is increased, and the oscillations correspondingly dephase, decreasing the observed amplitude.  These finite temperature effects give rise to a non-participating fraction $f_k$ of the atomic population drifting at $c^*$. Indeed, figure~5 shows that the majority of the ``thermal'' population surrounding the initial BEC is unaffected by the coupling.  Additionally, owing to imperfect preparation of the initial $\left|\uparrow,p_x=0\right>$ state, a fraction $f_0=0.15(10)$ remained at rest in $m_F = 0$ (and therefore did not participate at all in the -1 to +1 coupling).  Fluctuations in the background magnetic field also contribute to both $f_0$ and $f_k$.

The rest fraction $f_0$ was determined from long TOF images (such as those in figure~5). The drifting fraction $f_k$ was found by fitting a model
\begin{eqnarray}
\bar x (t,t_{\rm TOF}) = (1-f_0)\bigg\{(1-f_k)c^* &\bigg[ \frac{\sin(\Omega t+\phi_0)}{\Omega} \nonumber \\ &+ t_{\rm TOF} \cos(\Omega t+\phi_0)\bigg]+f_k c^*t + \bar x_0\bigg\},
\label{eq:fitfunc}
\end{eqnarray}
where $\bar x_0$ is the initial offset position, and $f_0$ is fixed at 0.15.  Using two sets of data, one in situ and one with $t_{\rm TOF} = 550~\upmu{\rm s}$, we performed joint fits for each laser intensity (four-photon coupling strength). In the initial analysis, we fix $c^*=2\hbar k_{\rm R}/m$ and fit the data to extract the parameters $f_k$, $\Omega$, $\phi_0$ and $\bar x_0$. The non-participating fractions for the data shown in figures~\ref{fig:zitter4}a,b are shown in figure~\ref{fig:zitter4}(c).  Next, we found the average of $f_k$ as a function of $\Omega$ to use in the model.  Finally, we remove the background slope due to the $f_k c^*t$ term  from the same five sets of data using a simple linear fit.  We refit the remaining signal to a ``fully participating'' model [(\ref{eq:fitfunc}) with $f_0 = f_k = 0$] with fixed $\Omega$ (from the original fit) to extract the effective speed of light parameter $c^*$.  We found the zitterbewegung amplitude (of the participating atoms) $c^*/\Omega$.   The model used to predict the amplitude is given by $(1-f_0)(1-f_k) 2\hbar k_{\rm R} /m\Omega$, and the uncertainty is dominated by our systematic uncertainty in $f_0$, which is due to magnetic field variations.   

The background slope due to the $f_k c^*t$ term was subtracted from the data presented in Figs.~2 and 3.  The curves are calculated from the values from the original fit using (\ref{eq:fitfunc}) without the $f_k c^*t$ term.  The average of the extracted $f_k$ was used in the theory curve in figure~\ref{fig:zitter4}(c) to show the expected amplitude of in situ oscillations.

\begin{figure}[h!]
\begin{center}
\includegraphics{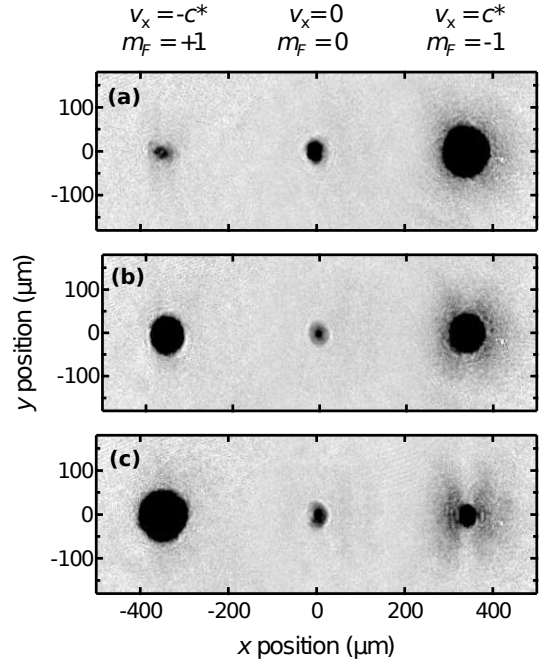}
\end{center}
\caption{Non-participating atoms.  These 30~ms TOF images, which reflect the momentum distribution of the atoms, were recorded at three evolution times: (a) $\Omega t \approx 0$, (b) $\Omega t \approx \pi/4$ and (c) $\Omega t \approx \pi/2$.  In (c), the transfer of the atoms from thermal cloud  was significant only near $p_x = 0$, leaving the majority of the thermal atoms behind, thus contributing to the non-participating fraction $f_k$.  These images show much larger BECs than were used for the in situ and short TOF measurements of Figs.~2 and 3, and have a correspondingly larger condensate fraction.  The atoms in the $m_F = 0$ state result from imperfect initial preparation, and contribute to the fraction $f_0$.}
\label{fig:zitter5}
\end{figure}

\section{Summary}
By engineering a two-level quantum system from initial states with opposite velocity, we reinterpret the ``curious'' physics of zitterbewegung in analogy to the Rabi oscillations ubiquitous in atomic physics.  In this language, the particles trembled because the initial state was not an eigenstate of the coupled system; once subject to the Dirac Hamiltonian, the system Rabi-oscillated between bare states of equal and opposite velocity.   As the atoms' coupling  was provided by resonant laser light instead of the electrons' rest energy, it is natural to think of a Rabi oscillation picture where the mass (coupling) is suddenly turned on and off. [Somewhat amusingly, the mechanism by which our laser field (a coherent state of light) generates mass is analogous to the Higgs mechanism where a Higgs condensate (a coherent matter wave) generates mass in the standard model~\cite{Bernstein1974}].  The zitterbewegung of electrons arises because two states -- particle and antiparticle states -- are coupled, and the resulting eigenstates are superpositions of the two.  Projections of bare electron states onto this basis result, as in the case of the atoms, in oscillations between states of opposite velocity. This straightforward analogy compels us to accept that the rest energy acts exactly as a coupling field   and mixes the particle and antiparticle states into eigenstates that are superpositions of the two.  

 While the Dirac equation generally applies only to fermionic systems in nature, quantum simulations such as ours directly realize Dirac-boson systems in the laboratory~\cite{Salger2011,Tarruell2012}, permitting access to new classes of experimental systems.  Though BECs near these Dirac points are short-lived~\cite{Qu2013,Spielman2009,Fu2011,Williams2012,ZhangPRA2013}, strong interactions, as are present near the superfluid-Mott transition in an optical lattice, can stably populate these states~\cite{Radic2012,Cole2012}, for example leading to bosonic composite-fermion states~\cite{Liu2009a,Sedrakyan2012}.

\ack
We thank V. Galitski, G.  Juzeli\=unas, and R.~F.~O'Connell for useful conversations.  This work was partially supported by the Office of Naval Research; by the Army Research Office with funds both from the Defense Advanced Research Projects Agency Optical Lattice Emulator program and the Atomtronics Multidisciplinary University Research Initiative; and by the NSF through the Physics Frontier Center at JQI. L.J.L. acknowledges the Natural Sciences and Engineering Research Council of Canada, K.J.-G. acknowledges CONACYT (Consejo Nacional de Ciencia y Tecnolog\'ia), and M.C.B. acknowledges NIST-American Recovery and Reinvestment Act.

\section*{References}

\providecommand{\newblock}{}

\end{document}